# Layer-by-Layer Dielectric Breakdown of Hexagonal Boron Nitride

*Yoshiaki Hattori*[*†], *Takashi Taniguchi*[‡], *Kenji Watanabe*[‡] *and Kosuke Nagashio*[**†§]

[†]Department of Materials Engineering, The University of Tokyo, Tokyo 113-8656, Japan
[‡]National Institute of Materials Science, Ibaraki 305-0044, Japan
[§]PRESTO, Japan Science and Technology Agency (JST), Tokyo 113-8656, Japan
[*]hattori@adam.t.u-tokyo.ac.jp, [**]nagashio@material.t.u-tokyo.ac.jp

**ABSTRACT:** Hexagonal boron nitride (BN) is widely used as a substrate and gate insulator for two-dimensional (2D) electronic devices. The studies on insulating properties and electrical reliability of BN itself, however, are quite limited. Here, we report a systematic investigation of the dielectric breakdown characteristics of BN using conductive atomic force microscopy. The electric field strength was found to be ~12 MV/cm, which is comparable to that of conventional $SiO_2$ oxides because of the covalent bonding nature of BN. After the hard dielectric breakdown, the BN fractured like a flower into equilateral triangle fragments. However, when the applied voltage was terminated precisely in the middle of the dielectric breakdown, the formation of a hole that did not penetrate to the bottom metal electrode was clearly observed. Subsequent *I-V* measurements of the hole indicated that the BN layer remaining in the hole was still electrically inactive. Based on these observations, layer-by-layer breakdown was confirmed for BN with regard to both physical fracture and electrical breakdown. Moreover, statistical analysis of the breakdown voltages using a Weibull plot suggested the anisotropic formation of defects. These results are unique to layered materials and unlike the behavior observed for conventional 3D amorphous oxides.

Since graphene was first successfully isolated from bulk graphite via mechanical exfoliation,[1] field-effect transistors (FETs) fabricated using many types of layered channel materials have attracted considerable attention.[2–4] Hexagonal boron nitride (BN), an insulating layered material with a wide band gap (5.2 – 5.9 eV),[5] is widely utilized as the substrate and gate insulator to achieve high carrier mobility in layered channel materials, especially graphene,[6] because its atomically flat surface with no dangling bonds considerably reduces its interaction with the graphene channel compared to that of a conventional $SiO_2$/Si surface. Moreover, the facile control of the number of layers of BN facilitates its use as a tunneling barrier in electronic devices.[7,8] Unlike the interfaces formed in conventional semiconductor devices using hetero-epitaxial techniques such as molecular beam epitaxy and pulsed laser deposition,[9] an interface formed by stacking BN with other layered materials is atomically definitive because of van der Waals interactions, and the intrinsic electrical structure of the channel material can be mostly retained.[10] Therefore, BN is considered to be an ideal insulator for layered channel materials. Several studies of the optical,[11] mechanical,[12] phonon,[13] electrical tunneling,[14,15] oxidation resistance,[16] and hydrophobic[17] properties of BN have also been reported.

The insulating properties and electrical reliability of the gate insulator are important issues in device applications. Although the dielectric breakdown voltage for BN has been recently reported,[14,15,18] little study has been conducted on the statistical analysis of the breakdown voltages and the breakdown mechanism. It is not clear whether BN is ideal from the viewpoint of the electrical reliability of the gate insulator. Thus far, the dielectric breakdowns of conventional three-dimensional (3D) amorphous oxides, such as the conventional $SiO_2$/Si system, have been extensively studied.[19–21] In general, the intrinsic electrical breakdown of an insulator can be simply explained by a percolation model,[22] in which defects that are generated locally under a high electrical field build up in the oxide until a critical defect density is reached, at which time the oxide suddenly and destructively breaks down. The variation in the threshold electrical field is characterized by the Weibull distribution,[23] which applies to weakest-link failure processes because dielectric breakdown is a stochastic process. The dielectric breakdown for the high-$k$ oxides on Si[24] and graphene[25] also follows the percolation model. However, it has not been determined whether the dielectric breakdowns of 2D layered materials follow the conventional percolation model for 3D amorphous oxides. In this study, the dielectric breakdown behavior of BN was systematically investigated using conductive atomic force microscopy (C-AFM) to reveal the breakdown behavior of 2D layered materials.

**RESULTS AND DISCUSSION**

In this study, high-quality BN grown using a temperature-gradient method under a high-pressure and high-temperature atmosphere[5] was used to obtain reliable data. Although other growth techniques are available,[18,26–28] their quality still requires improvement. The BN was transferred to a silicon wafer coated with a 50-nm layer of platinum using the mechanical exfoliation technique. A schematic diagram of the experimental setup is presented in **Figure 1a**. A rhodium-coated silicon cantilever with a radius of curvature of ~100 nm was used because the high melting point ($T_M$=2236 K) and mechanical hardness of rhodium enable the iterative usage of the cantilever. A protective resistor ($R_{prot}$) of 100 MΩ was inserted into the circuit in series to protect the coating of the AFM tip from being subjected to an abrupt, high current at breakdown.[29] This $R_{prot}$ had no effect on the precise measurement of the leakage current because it was much smaller than the resistance of BN ($R_{BN}$). The current ($I$) - voltage ($V$) measurement was performed at room temperature in ambient air using a semiconductor device parameter analyzer that was assembled externally to the AFM system. The positive bias voltage with respect to the grounded AFM tip was applied to the Pt electrode and gradually increased to 40 V in a staircase sweep measurement (time-zero test). Although we have examined the positive bias as well as negative bias for the breakdown, the polarity effect is not so obvious. Therefore, the positive bias was selected in this study. The typical values of the voltage step, ramping rate, and integral time were 0.050 V, 0.89 V/s, and 40 ms, respectively.

**Figure 1b** illustrates the electric field around the AFM tip, which was calculated using a simplified model (details provided in

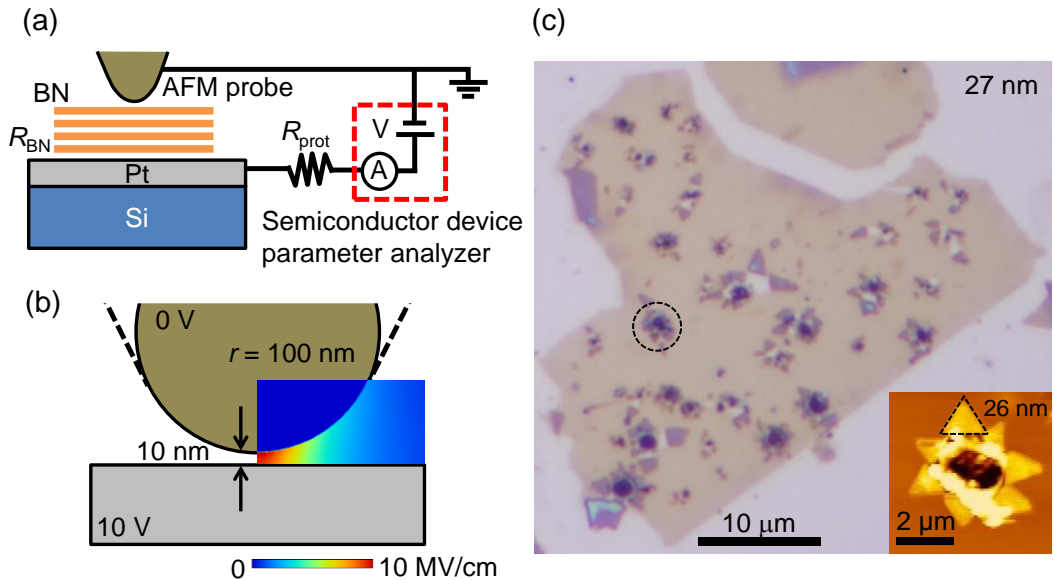

**Figure 1**: (a) Schematic diagram of the measurement setup using the C-AFM system. (b) Intensity profile of the electric field around the AFM probe with a 10-V bias applied. (c) Optical microscopy image of a 27-nm-thick BN flake after a series of breakdown tests. One of the fractured areas is highlighted with the broken circle. The inset presents a typical AFM topographical image of a fractured fragment.

Supplemental **Figure S1**). The area of ~40 nm in diameter below the AFM tip can be treated approximately as a parallel-plate electrode, which is consistent with a previous estimation.[30] It is well known that the surface of $SiO_2$ is modified during a C-AFM measurement because of the electrochemical reactions catalyzed by the adsorbed water resulting from anodic oxidation.[29,31,32] It is, however, expected that the effect of adsorbed water on electrical measurements may be negligible for BN because BN is hydrophobic[17] and resistant to oxidation.[16] Indeed, our supplemental experiment revealed no evidence of surface modification in BN, unlike $SiO_2$/Si (details provided in Supplemental **Figure S2**).

*I-V* measurements of local areas (*i.e.*, breakdown tests) were performed repeatedly on different areas of the same BN flake. **Figure 1c** presents an optical microscopy image of a BN flake with a 27-nm thickness after the breakdown test. The dielectric breakdown led to the catastrophic fracture of the BN. The mechanical stress that was abruptly released during the breakdown event resulted in fracture propagation. The fractured fragments often exhibited equilateral triangular shapes, especially for thick BN flakes, as shown in the inset of **Figure 1c**. The thickness of the triangular fragments was equal to that of the original flake. This observation strongly suggests that the preferred fracture directions could be armchair and/or zigzag directions, as expected from the hexagonal lattice structure of BN.[33,34] These unique breakdowns can be also seen in the negative bias voltage stress.

**Figure 2a** presents three sets of *I-V* curves for the same BN flake with a thickness of 11 nm. The current increased smoothly with increasing voltage up to approximately 15 V and then increased suddenly at the voltages indicated by solid arrows. In most cases, the current decreased again within a short time, forming sharp peaks. Then, the current fluctuated randomly until the conduction path had completely formed between the two electrodes, where the current was controlled by $R_{prot}$. This variation in the voltages, indicated by the dotted arrows, was clearly larger than the initial rapid increases in voltage (solid arrows). This finding suggests that the dielectric breakdown character of BN is reflected in these initial rapid increases.

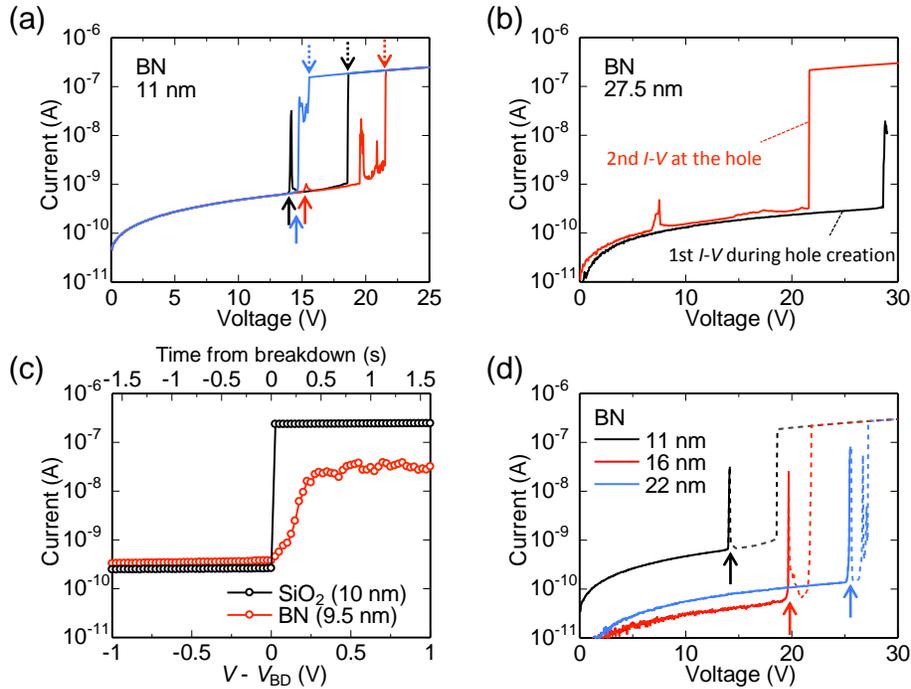

**Figure 2**: (a) Three *I-V* curves for the same BN flake with a thickness of 11 nm. The solid and dotted arrows indicate the $V_{BD}$ values and the short-circuit voltages, respectively. (b) *I-V* curves for a fresh BN flake (first *I-V* curve) and the BN layers remaining under the hole formed during the first measurement (second *I-V* curve). (c) Comparison of *I-V* curves for $SiO_2$ and BN, where the transverse axis represents $V - V_{BD}$. The voltage step, ramping rate, and integral time for both experiments were 0.025 V, 0.63 V/s and 20 ms, respectively. (d) *I-V* curves for BN samples with different thicknesses. The current curves after the initial peaks are represented by dotted lines because they do not represent the dielectric breakdown of BN samples.

These initial points of rapid increase were investigated in detail. A simplified *I-V* curve is illustrated in **Figure 3a**. To observe the topographical changes before and after the initial rapid increase, the applied voltage was intentionally terminated at three different target points. Because the time duration of the initial current increase was relatively long, the voltage could be successfully terminated at the target points. Although there was no detectable change in the AFM topographical image at (i), the formation of a hole with a diameter of ~300 nm and a depth of ~0.6 nm was observed on the BN surface at (ii), as shown in **Figure 3b**. The hole did not reach the bottom Pt electrode at (ii), but the subsequent catastrophic fracture resulted in a large hole at (iii), as shown in **Figure 3c**. The AFM topographical image revealed that both the Pt coating and the BN flake were broken, which was confirmed by the detection of the Si wafer underneath the Pt coating in Raman measurements (details provided in Supplemental **Figure S3**).

Now, our greatest interest is in whether the remaining BN layers beneath the hole are still electrically inactive. In the case of thicker BN, the fracturing could be more easily stopped in the middle of the BN, as shown in **Figure 3d**. After the AFM tip was adjusted to precisely the center of the hole, *I-V* measurements were performed, and the results are presented in **Figure 2b**. In the low-voltage region, the current was slightly higher than that measured in the first *I-V* curve corresponding to the creation of the hole, but the difference was minimal. Then, a peak appeared, suggesting the onset of dielectric breakdown. Therefore, it was clear that the remaining BN layers beneath the hole had not yet broken down electrically.

Based on these observations, it appears that BN undergoes layer-by-layer breakdown in terms of both physical fracture and electrical breakdown.

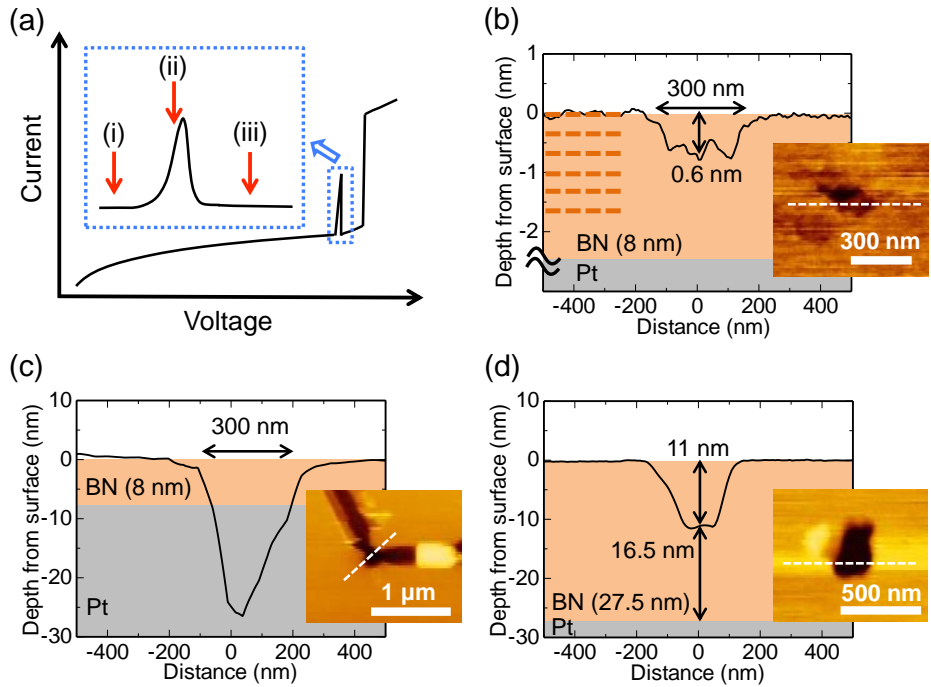

**Figure 3**: (a) Schematic representation of the *I-V* curve for BN. The applied voltage was intentionally terminated at three different points, (i–iii). (b),(c) Height profiles and AFM topographical images for BN breakdown terminated at (ii) and (iii), respectively. (d) Height profile and AFM topographical image for thicker BN. Although the AFM topographical image was acquired again using an AFM tip with a smaller radius of ~10 nm, the hole shape did not change. Additionally, no additional small hole was detected within the fractured large hole, suggesting that the BN was not completely physically penetrated.

Figure 4 presents a series of schematic illustrations of the breakdown process. The bonding energy along the direction vertical to the BN surface will vary periodically because of $sp^2$ covalent bonds in the layer and van der Waals bonds in the interlayer, as shown in **Figure 4a**. The breakdown begins in the top layer because of its 2D nature and proceeds layer by layer toward the bottom Pt electrode, as shown in **Figures 4b** and **4c**, respectively. Then, the decrease in the BN thickness beneath the AFM tip enhances the electric field, resulting in sudden catastrophic fracture (**Figure 4d**). The AFM tip is often lifted up into the neighboring region because of the resulting mechanical shock, as shown in **Figure 4e**. This causes the current to decrease sharply, which leads to the formation of the initial sharp peaks observed in **Figures 2a** and **3a**. In fact, when an AFM image was acquired after breakdown, the position of the image was sometimes shifted by a few micrometers with respect to the original image, where the displacement of the tip was larger than the usual drift. Finally, similar behavior repeats until complete conduction is achieved between the two electrodes. Therefore, the initial peak indicates the dielectric breakdown, and the complete conduction observed later is an artificial effect. Hereafter, the voltage corresponding to the onset of the initial peak is defined as the dielectric breakdown voltage ($V_{BD}$).

From the investigation of the initial current increase, we realized that the increase in the current for the BN at dielectric breakdown was much slower than that for the well-studied case of $SiO_2$ oxide. Therefore, the dielectric breakdown for $SiO_2$ with the same thickness as the BN was also tested using the same C-AFM system for comparison. **Figure 2c** presents a comparison of the $I$-$V$ curves for BN and $SiO_2$ near the dielectric breakdown voltage, where the transverse axis represents $V - V_{BD}$. For 4 experiments, the time duration of the rapid current increase for $SiO_2$ was less than the minimum hold time of 20 ms for the $I$-$V$ measurement system. On the other hand, the average time duration and the standard deviation of the same BN flake for 15 experiments were 253 ms and 87 ms, respectively. The origin of this difference may be the layer-by-layer breakdown behavior of BN, which seems to be the primary difference in behavior between 2D layered materials and 3D amorphous oxides.

Here, let us discuss the dielectric breakdown behavior of BN from the microscopic point of view. Avalanche breakdown is one of the mechanisms of dielectric breakdown.[35] The carriers that are accelerated by the electrical field gain energy greater than the band gap to excite electron-hole pairs, which often breaks the bonding. For BN, the generation of electron-hole pairs might occur only within each layer because the interlayer distance is 2.3 times greater than the distance between nearest-neighboring atoms in a layer. If this is the case, then the observed layer-by-layer breakdown characteristics are not specific to C-AFM measurements but rather are intrinsic to layered materials, although in the case of a conventional metal top electrode, the first layer to break may not be limited to the top layer as it is in the case of an AFM tip.

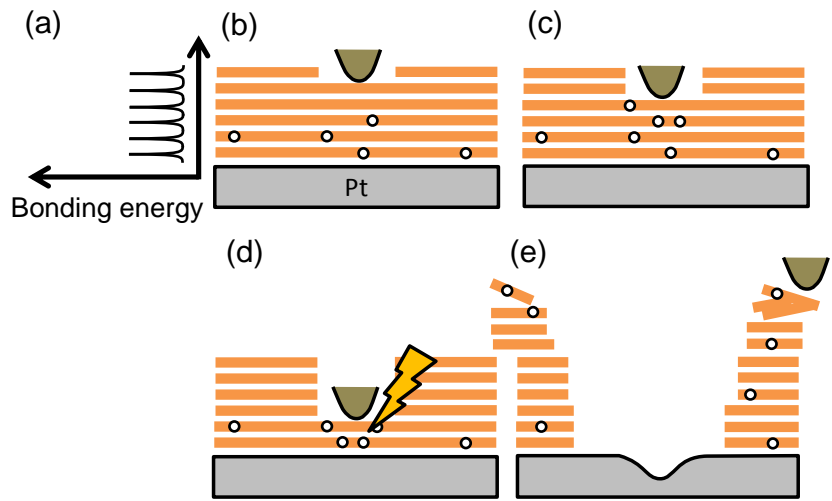

**Figure 4**: A series of schematic illustrations of the layer-by-layer breakdown process. (a) Bonding energy along the direction vertical to the layer surface. The breakdown begins in the top layer (b) and proceeds layer by layer (c). The enhancement of the electric field caused by the thickness reduction (d) induces sudden catastrophic fracture (e). Irreversible defects are drawn as open circles.

Let us consider the formation of irreversible defects, that is, the soft breakdown before the layer-by-layer hard breakdown. Iterative $I$-$V$ measurements for the BN flake with a thickness of 27.5 nm were performed from 0 V to 25 V (a few volts below $V_{BD}$). All measurements clearly overlapped (details provided in Supplemental **Figure S4**), indicating no detectable irreversible defect formation. Now, let us move back to the case of the dielectric breakdown for the remaining BN layer under the hole, as shown in **Figure 2b**. The $V_{BD}$ value for the remaining 16.5-nm BN layer under the hole was ~6.7 V. This is much lower than that for a fresh BN flake with the same thickness ($V_{BD}$ = ~20 V for ~16.5 nm) because of the irreversible defect formation in the remaining BN layer under the hole. Thus, it was concluded that irreversible defects must form immediately just prior to dielectric breakdown. Therefore, these irreversible defects are schematically drawn as open circles in **Figure 4**.

The variation in $V_{BD}$ for different thicknesses was investigated to gain further insight into the difference between 2D layered materials and 3D amorphous oxides. **Figure 2d** presents typical $I$-$V$ curves for BN flakes with different thicknesses. $V_{BD}$, as indicated by the arrows, increased with increasing thickness. The measured $V_{BD}$ values were sorted in ascending order for statistical analysis, and then the cumulative failure probability ($F_{BD}$) against $V_{BD}$ was characterized by Weibull plots.[36] Weibull plots of the $V_{BD}$ distributions for different thickness are presented in **Figure 5a**. The correlation coefficient values of the linear fits for BN with thicknesses of 11 nm, 16 nm, and 22 nm are 0.983 (for 15 samples, N = 15), 0.965 (N = 17), and 0.945 (N = 16), respectively. The good linear fits indicate that the $V_{BD}$ data can be described using the Weibull plot parameters. Each fit line is described by the following equation,

$$\ln(-\ln(1 - F_{BD})) = \beta \ln(V_{BD}) - \beta \ln \alpha,$$

where $\alpha$ corresponds to the $V_{BD}$ at which 63.2% of samples fail, and $\beta$ is the slope of the line, which characterizes the degree of $V_{BD}$ variation (uniformity) and determines the failure mode classification that is widely used in reliability engineering. In general, $\beta < 1$ indicates extrinsic breakdown caused by extrinsic defects (B-mode breakdown) as well as early failure (A-mode breakdown).[25,36] Therefore, $\beta > 1$ (C-mode breakdown) is required in terms of reliability because this mode represents the intrinsic nature of the dielectric. In this study, the $\beta$ values for 11-nm, 16-nm, and 22-nm BN were estimated to be 23, 19, and 13, respectively, reflecting intrinsic breakdown. By comparison with the value of $\beta = \sim 8.0$ observed for high-$k$,[37] the quality of the present BN crystal is found to be reasonably high, suggesting that the intrinsic breakdown mechanism can be discussed based on these data. As seen in **Figure 5a**, thicker BN samples had smaller $\beta$ values in the tested range. Interestingly, this result is opposite to that observed in the SiO$_2$ case, which is well explained by the percolation model.[22] For thicker oxides, many defects are required to create a breakdown path. When the randomness of defect formation in 3D amorphous oxides and the lack of interaction between defects are taken into account, the variation in $V_{BD}$ becomes small. This is the case for SiO$_2$. The opposite results observed for 2D layered BN suggest that the anisotropic formation of defects is the key to understanding the present data; further analysis via simulation is required to investigate this issue.

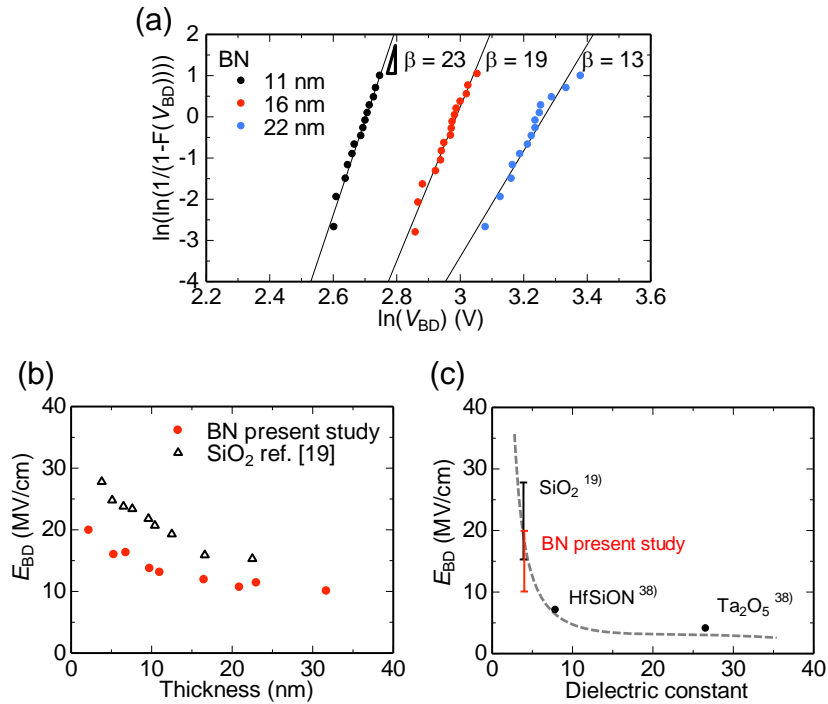

**Figure 5**: (a) Weibull plots of $V_{BD}$ distributions for different thicknesses of BN. (b) $E_{BD}$ as a function of thickness for BN for the data measured in this study and for previously reported data. (c) The relation between $E_{BD}$ and dielectric constant $k$.

**Figure 5b** presents the electric field strength ($E_{BD} = V_{BD}$ / thickness) as a function of thickness for BN based on the data measured in this study as well as previously reported data for SiO$_2$.[19] Each point represents data averaged from more than 5 measurements of the same flake. The $E_{BD}$ obtained in this study is in agreement with that reported previously for thin BN.[15] It is interesting that $E_{BD}$ for BN decreases with increasing the thickness. The similar behavior is also observed for SiO$_2$. This could be explained as follows. Because of the very high density of defects formed in thicker BN flakes, the local internal fields may well reach the magnitudes observed in thin BN flakes. This also explains the lower $E_{BD}$ values observed for slower ramping rates (details presented in Supplemental **Figure S5**) because more defects are formed in such cases. Although the increase in atomic bonding energy from bulk BN to monolayer BN[13] may contribute to the increase in $E_{BD}$ observed with decreasing thickness, it does not explain the difference in $E_{BD}$ observed for different ramping rates.

Finally, let us discuss the relation between $E_{BD}$ and the dielectric constant ($k$), which is important in terms of the selection of materials for use as gate insulators in FETs. Empirically, $E_{BD}$ and $k$ exhibit an inverse relationship,[37–39] as represented by the dotted line in **Figure 5c**. Materials with ionic bonding exhibit larger $k$ values because of their higher polarization, whereas those with covalent bonding exhibit smaller $k$ values. The $E_{BD}$ values obtained for the present BN flakes are plotted in **Figure 5c** under the assumption that $k = 4$. The data lie along the empirical relation. For $E_{BD}$ values obtained by applying a voltage normal to the layer surface, the anisotropic nature of the covalent bonding in BN is not apparent. $E_{BD}$ values obtained by applying the voltage parallel to the layer surface are expected to differ considerably.

**CONCLUSION**

The dielectric breakdown of high-quality BN flakes was investigated via C-AFM to reveal the breakdown behavior of 2D layered materials. Based on detailed $I$-$V$ measurements of the BN layers remaining after hole formation, layer-by-layer breakdown was confirmed in the context of both physical fracture and electrical breakdown. Statistical analysis of the breakdown voltages using Weibull plots suggested the anisotropic formation of defects. These results are unique to layered materials and differ from those observed for conventional 3D amorphous oxides. Moreover, the robustness of the irreversible defect formation and the high electric field strength suggest that BN is a reliable insulator for use in 2D electronic devices.

**EXPERIMENTAL SECTION**

Single-crystal BN flakes exfoliated on a Pt/Si substrate were identified using an optical microscope (LV100, Nikon) equipped with a camera (DP71, Olympus). A platinum electrode was selected because it is stable in ambient air, without any native oxide formation, and the root-mean-square (RMS) surface roughness is less than 0.2 nm. The thicknesses of the BN flakes (2 – 30 nm) were evaluated via AFM in contact mode. For comparison, $SiO_2$ films grown thermally on n-type Si (100) wafers were also investigated in the same system with the protective resistor of 100 MΩ. The thickness of the $SiO_2$ was measured using an ellipsometer (GES 5, Sopra). Topographic images were acquired using an AFM (a SPA-400 microscope unit controlled by an SPI 400 probe station, SII NanoTechnology) with a rhodium-coated Si cantilever (SI-DF20-R, SII NanoTechnology) with a loading force of 1.6 – 1.9 nN at room temperature in ambient air. A stiffness of the cantilever is 1.6 – 1.9 N/m. *I-V* measurements were performed using a semiconductor device parameter analyzer (B1500A, Keysight Technology) assembled externally to the AFM system.

## SUPPORTING INFORMATION

Details of the numerical calculation of the electrostatic field around the AFM tip; Comparison of AFM images of $SiO_2$ and BN acquired at a bias voltage of 15 V; Raman spectra of BN after dielectric breakdown; Iterative *I-V* measurements; $V_{BD}$ as a function of ramping rate. This material is available free of charge via the Internet at http://pubs.acs.org.

## AUTHOR INFORMATION

**Corresponding Author**
Email:*hattori@adam.t.u-tokyo.ac.jp,
**nagashio@material.t.u-tokyo.ac.jp
**Notes**
The authors declare no competing financial interests.

## ACKNOWLEDGMENTS

This research was partly supported by Grants-in-Aid for Scientific Research on Innovative Areas and for Research Activity Start-up by the Ministry of Education, Culture, Sports, Science and Technology of Japan.

**Supplemental note:**

# Layer-by-Layer Dielectric Breakdown of Hexagonal Boron Nitride

*Yoshiaki Hattori*[*†], *Takashi Taniguchi*[‡], *Kenji Watanabe*[‡] *and Kosuke Nagashio*[**†§]

[†]Department of Materials Engineering, The University of Tokyo, Tokyo 113-8656, Japan

[‡]National Institute of Materials Science, Ibaraki 305-0044, Japan

[§]PRESTO, Japan Science and Technology Agency (JST), Tokyo 113-8656, Japan

[*]hattori@adam.t.u-tokyo.ac.jp, [**]nagashio@material.t.u-tokyo.ac.jp

**Numerical calculation of the electrical field.**

The electrostatic field around the AFM tip under the application of a bias voltage was calculated. The potential ($\phi$) in the space around the AFM tip, with the exception of the inside of the electrode, satisfies Poisson's equation:

$$\text{div}\,(\varepsilon\,\text{grad}\,\phi) = 0$$

where $\varepsilon$ is the dielectric constant. The electrostatic field ($E$) is calculated as $-\text{grad}\,\phi$. The governing equation can be solved explicitly using the finite-volume method in a simplified 2D model (**Figure S1a–b**) with a cylindrically symmetric (r, z) geometry. The calculation area was divided into 256×256 equally sized grids. The boundary conditions were defined to be adiabatic along the right lateral surface and top surface and to be isothermal at the bottom at a constant applied bias voltage. The potential inside the AFM tip was set to zero. The calculation results are presented in Supplemental **Figure S1c–f**.

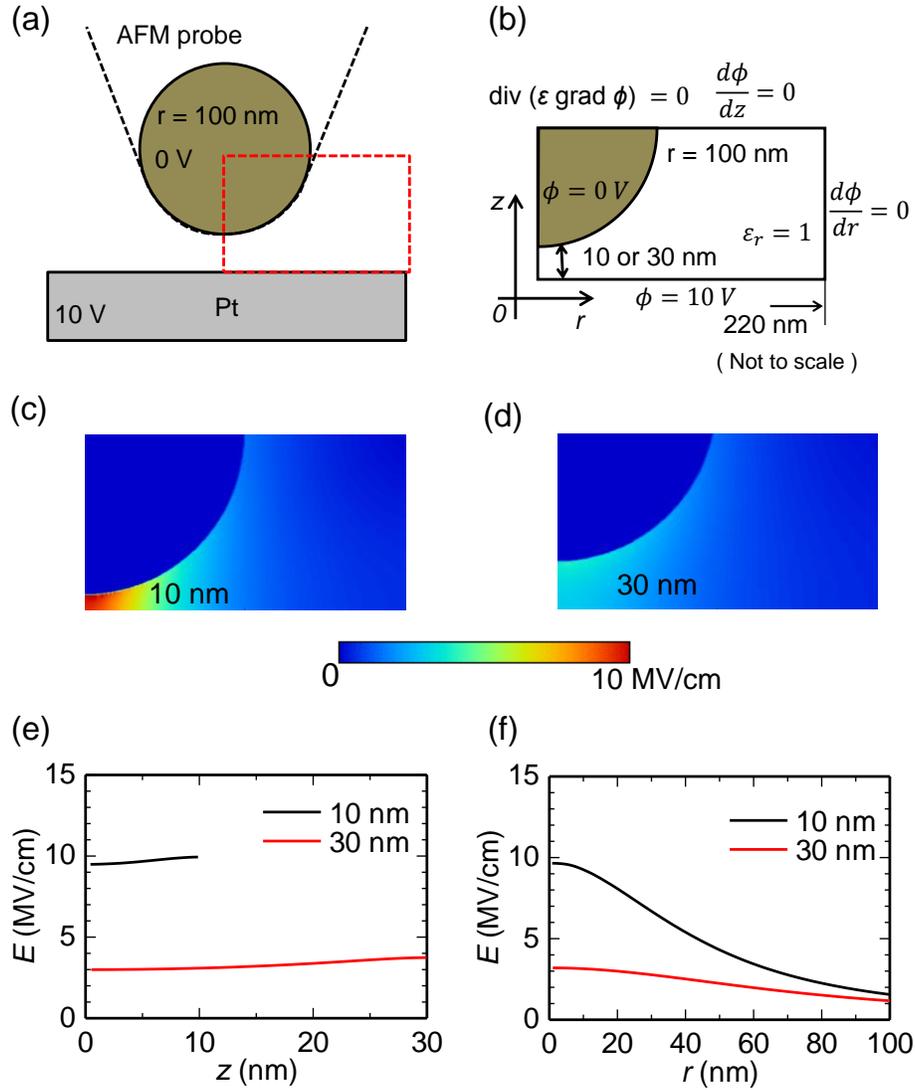

**Supplemental Figure S1:** 2D simulation model (a–b). Electric field around the AFM tip for gap sizes of 10 nm (c) and 30 nm (d). The color scale indicates the electric field distribution calculated by solving Poisson's equation. Enlarged views of the electric field at the center of the AFM tip along the z direction (e) and at the midpoint between the AFM tip and the Pt surface along the r direction (f).

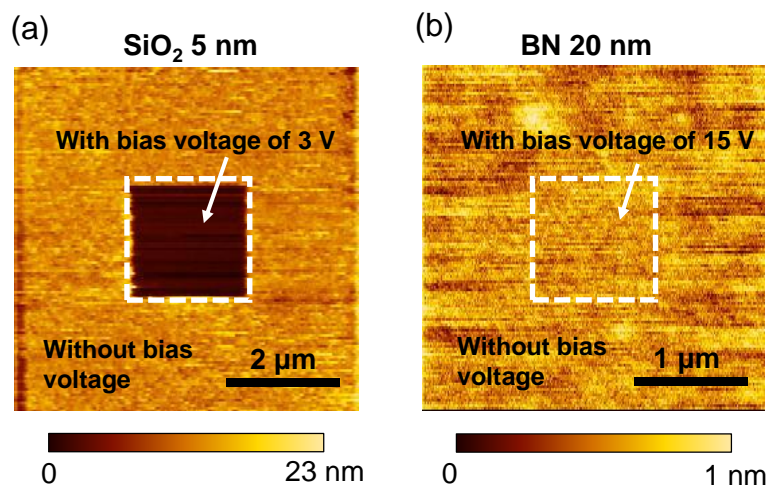

**Supplemental Figure S2:** AFM topographical images of SiO$_2$ (a) and BN (b). The areas scanned with bias voltages are indicated by dotted squares. The scan frequency was 1 Hz. A noticeable change in height was observed for the SiO$_2$, whereas the atomically flat surface of the BN, with an RMS of 0.14 nm, was maintained even after scans at an applied voltage of 15 V.

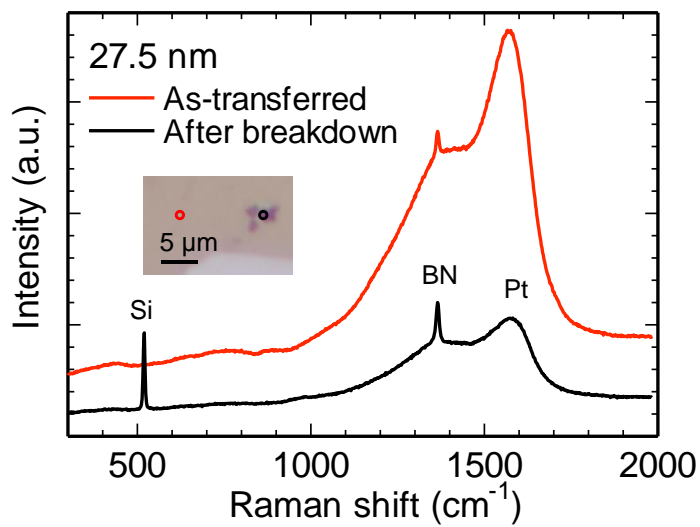

**Supplemental Figure S3:** Raman spectra of 27.5-nm BN, both as transferred and after breakdown, acquired using an Ar laser with $\lambda$ = 488 nm. The Raman E$_{2g}$ peak for BN was observed at 1335 cm$^{-1}$. The broad peak near 1576 cm$^{-1}$ was attributed to the Pt layer on the Si wafer. After the hard dielectric breakdown, a peak associated with the Si substrate appeared at 520 cm$^{-1}$, indicating that both the Pt layer and the BN flake were fractured upon the hard dielectric breakdown.

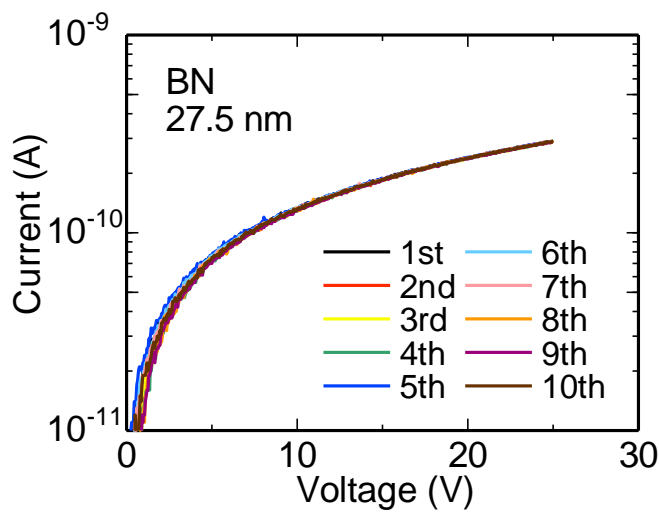

**Supplemental Figure S4:** Ten iterative *I-V* measurements from 0 V to 25 V acquired at the same position on BN with a thickness of 27.5 nm. The $V_{BD}$ values at different positions for the same BN flake fall in the range of 27 – 30 V.

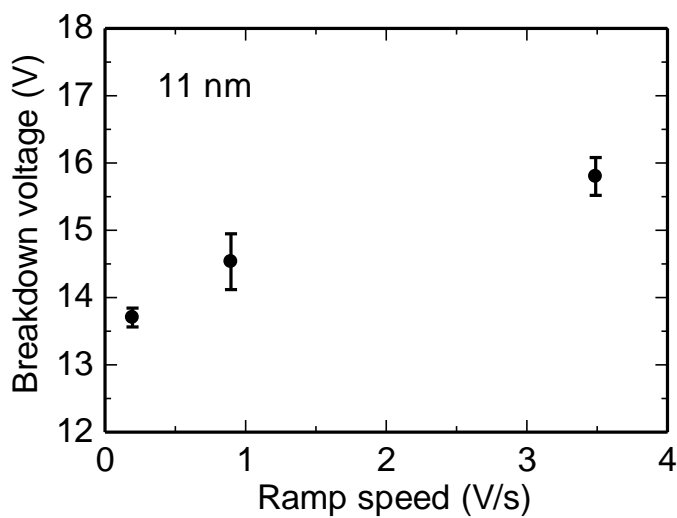

**Supplemental Figure S5:** Breakdown voltage as a function of ramping rate. The breakdown tests were performed on the same BN flake, with a thickness of 11 nm. Solid circles represent average values determined from more than 5 breakdown tests.